\documentclass[graybox]{svmult}

\usepackage{mathptmx}       
\usepackage{amsmath, mathtools,amssymb}  
\usepackage{helvet}         
\usepackage{courier}        
\usepackage{type1cm}        
 \usepackage{caption}                           
\usepackage[]{natbib}                         
\usepackage{makeidx}         
\usepackage{graphicx}        
\usepackage{float}

\usepackage{subcaption}
\usepackage{multicol}        
\usepackage[bottom]{footmisc}

\newtheorem{exmp}{Example}[section]

\makeindex             

\begin{document}

\title*{Semi-parametric time series modelling with autocopulas}
\author{Antony Ware, Ilnaz Asadzadeh}
\institute{A. Ware \at University of Calgary, 2500 University Drive NW, Calgary, AB T2N 1N4 , \email{aware@ucalgary.ca}
\and I. Asadzadeh \at University of Calgary, 2500 University Drive NW, Calgary, AB T2N 1N4 \email{iasadzad@ucalgary.ca}}
%
%
\maketitle

\abstract{In this paper we present an application of the use of autocopulas for modelling financial time series showing serial dependencies that are not necessarily linear. 
The approach presented here is semi-parametric in that it is characterized by a non-parametric autocopula and parametric marginals. One advantage of using autocopulas is that they provide a general representation of the auto-dependency of the time series, in particular making it possible to study the interdependence of values of the series at different extremes separately. The specific time series that is studied here comes from  daily cash flows involving the product of daily natural gas price and daily temperature deviations from normal levels. Seasonality is captured by using a time dependent normal inverse Gaussian (NIG) distribution fitted to the raw values.}

\section{Introduction}
\label{sec:1}

In this study, autocopulas are used to characterise the joint distribution between successive observations of a scalar Markov chain. A copula joins a multivariate distribution to its marginals, and its existence is guaranteed by Sklar's theorem \cite{sklar1959fonctions}. In particular, a Markov chain of first order with any given univariate margin can be constructed from a bivariate copula. 
A theoretical framework for the use of copulas for simulating time series was given by \cite{darsow1992copulas}, who presented necessary and sufficient conditions for a copula-based time series to be a Markov process, but not necessarily a stationary one. They presented theorems specifying when time series generated using time varying marginal distributions and copulas are Markov processes. \cite{joe1997multivariate} proposed a class of parametric stationary Markov models based on parametric copulas and parametric marginal distributions. \cite{chen2006estimation} studied the estimation of semiparametric stationary Markov models, using non-parametric marginal distributions with parametric copulas to generate stationary Markov processes. 

The term 'autocopula' was first used to describe the unit lag self dependence structure of a univariate time series in \cite{rakonczai2012autocopulas}, and we adopt the terminology here. We make use of the framework presented in \cite{darsow1992copulas} to produce Markov processes such that the marginal distribution changes over time.  The main benefit of using autocopulas for univariate time series modelling is that the researcher is able to specify the unconditional (marginal) distribution of $X_{t}$ separately from the time series dependence of $X_{t}$ (\cite{patton2009copula}).  We apply a semiparametric method which is characterized by an empirical autocopula and a parametric time varying marginal distribution. This allows us to capture seasonal variations in a natural way. This is an important feature of our model, motivated by the fact that many financial and economic time series exhibit seasonality, particularly those arising from energy and commodity markets.
The remainder of this paper is organized as follows. In Section~\ref{sec:2}, we introduce the data; Section~\ref{sec:3} describes the model, including a review of copulas, and of the Normal Inverse Gaussian distribution. This section also includes details of the calibration and simulation procedures, and the final section presents some results.


\section{The data}
\label{sec:2}
The motivation from this project came from the desire to develop a parsimonious model that could capture so-called load-following (or swing) risk. This is one of the main sources of financial uncertainty for an energy retailer, and arises from the combination of retail customer consumption (volume) uncertainty and price uncertainty. Both volume ($V$) and price ($P$) are driven to a large extent by weather. In particular, average daily temperature is one of the main drivers of daily natural gas consumption in various North American markets: this in turn drives market prices through a supply and demand process.

Some of the load-following risk exposure can be hedged using gas forwards and temperature derivatives. The most significant part that cannot be easily hedged is directly linked to the daily product between the weather deviation from normal and the daily price deviation from the expected value of the ex-ante forward price. To make this more specific, let $\overline{P}$ denote the last-traded forward monthly index price, and $\overline{V}$ the expected monthly average volume. Cash flows for the retailer depend on the product $PV$, and the uncertainty in this quantity can be written 
\[ (\overline{P}+\Delta P)(\overline{V}+\Delta V)-\overline{P}\overline{V} = \overline{P}\Delta V+\overline{V}\Delta P+\Delta P \Delta V. \]
We posit a linear relationship between volume and weather deviations, so that $ \Delta V = \beta   \Delta W + \epsilon$, where $\beta$ is the sensitively of consumption to weather, which can be determined from load data for different regions. Forward instruments in weather and natural gas markets can then be used to hedge risks corresponding to the terms $\overline{P}\Delta V$ and $\overline{V}\Delta P$. Apart from the error term $\epsilon$ in the volume-weather relationship, which we assume to be relatively small, it can be seen that the term $\Delta P \Delta W$ becomes the main driver of unhedged risk in these cashflows. One approach would be to develop separate models for weather and natural gas prices (both daily and forward prices). However, because of the desire for parsimony, we instead seek a model that allows us to study the time-series $X_t=(\Delta P \Delta W)_t$, in order to estimate the range and probabilities of possible outcomes at the level of a complex portfolio of retail load obligations. 

\begin{figure}[H]
\sidecaption[t]
\includegraphics[width=0.62\linewidth]{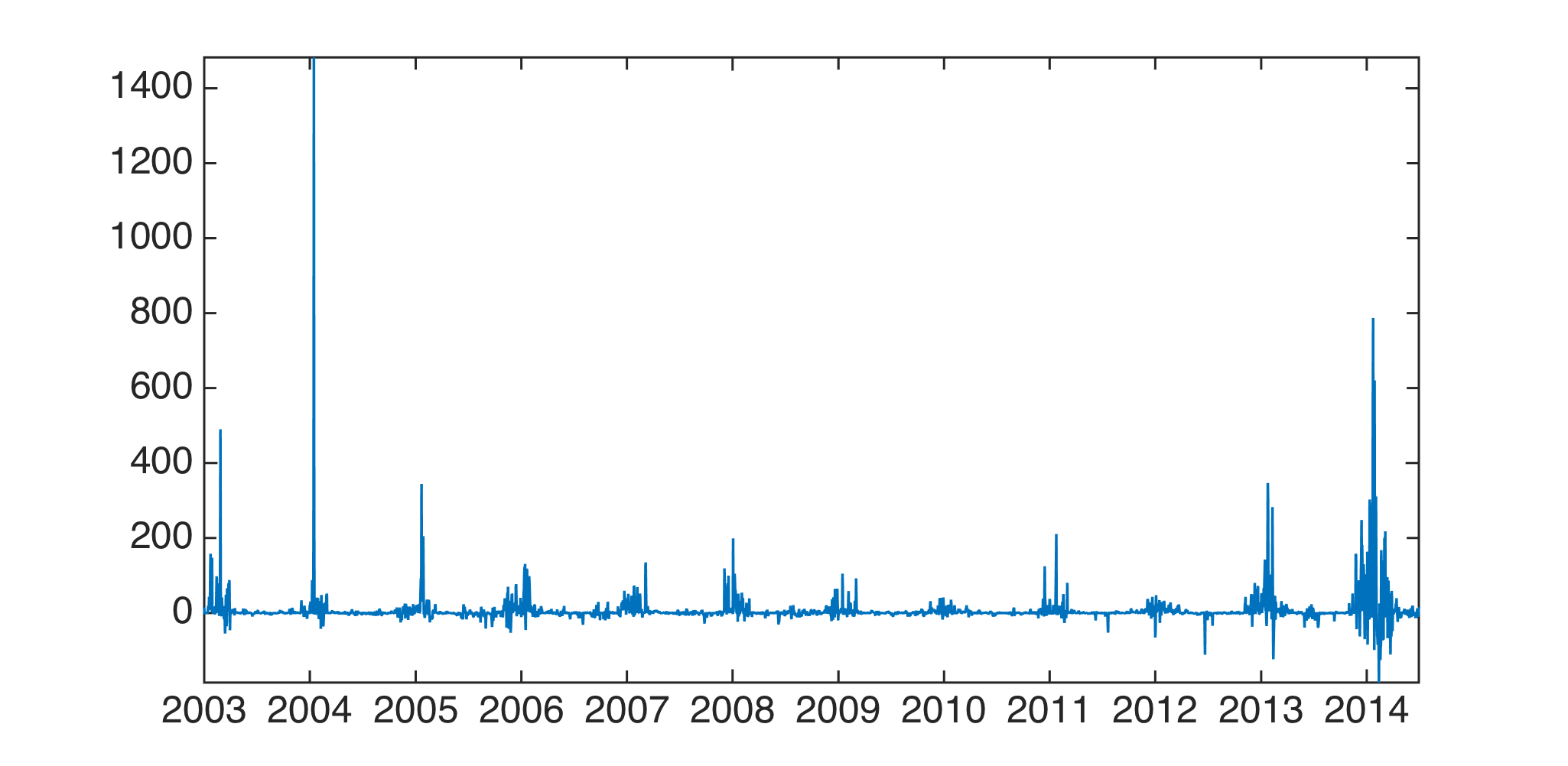}
\caption{Product of weather and gas price deviations ($\Delta P\Delta W$) in Algonquin over 2003-14. Spikes correspond to combinations of high weather deviation from normal and high spot price deviation from next forward month.}
\label{fig:one}
\end{figure}

Here we study the North American market and focus on the Algonquin location for the weather data. The data cover the period 1 January 2003 - 31 June 2014 on a daily basis, and are shown in Figure~\ref{fig:one}. The most dramatic feature of the graph is the presence of intermittent clusters of spikes, during which the gas prices rise from their approximate average daily value and at the same time temperature rises or falls drastically. These mostly occur during winter, although large deviations also occur at other times of the year. It is also clear that the marginal densities of these observations will not be well-represented by normal distributions.

\section{The model}
\label{sec:3}
Here we introduce the simulation model in more detail, providing a brief review of copulas, and the normal inverse Gaussian distribution, which we use for the marginal densities.
\subsection{Copulas and autocopulas}
\label{subsec:copulas}

A copula\footnote{For more discussion on the theory of copulas and specific examples, see \cite{nelsen2007introduction}.} is a multivariate distribution function defined on a unit cube $[0,1]^{n}$, with uniformly distributed marginals. In the following, we use copulas for the interdependence structure of time series and, for simplicity and the fact that we are interested in the first order lag interdependence, we focus on the bivariate case, although the approach can be used to capture dependence on higher order lags.

Let $F_{12}(x,y)$ be the joint distribution function of random variables $X$ and $Y$ whose marginal distribution functions, denoted as $F_{1}$ and $F_{2}$ respectively, are continuous. Sklar's theorem specifies that there exists a unique copula function $C(u,v) = F_{12}(F_{1}^{-1}(u), F_{2}^{-1}(v))$ that connects $F_{12}(x,y)$ to $F_{1}(x)$ and $F_{2}(y)$ via $F_{12}(x,y) = C(F_{1}(x),F_{2}(y))$. The information in the joint distribution $F_{12}(x,y)$ is decomposed into that in the marginal distributions and that in the copula function, where the copula captures the dependence structure between $X$ and $Y$. Various families of parametric copulas are widely used (Gaussian, Clayton, Joe, Gumbel copulas, for example). 
  
In a time series setting, we use a copula (or \emph{autocopula}) to capture the dependence structure between successive observations.  More generally, we have the following definition (\cite{rakonczai2012autocopulas}).

\begin{definition}[Autocopula]
Given a time series $X_{t}$ and $\mathcal{L}= \{ l_{i} \in \mathbb{Z}^{+}, i = 1,...,d\}$ a set of lags, the autocopula $C_{X,\mathcal{L}}$ is defined as the copula of the $d+1$ dimensional random vector $(X_{t},X_{t-l_{1}},...,X_{t-l_{d}})$.
\end{definition}

If a times series $X_t$ is modelled with an autocopula model with unit lag, with autocopula function $C(u,v) = C_{X,1}(u,v)$, and (time-dependent) marginal CDF $F_t(x)$, then, for each $t$,  the CDF of the conditional density of $X_{t}$ given $X_{t-1}$ can be expressed
\begin{equation}
\label{eq:conditionalCDF} 
F_{X_t|X_{t-1}}(x) = \frac{\partial C}{\partial u}\big(F_{t-1}(X_{t-1}),F_t(x)\big).
\end{equation}
We will discuss issues related to calibration and simulation below.

Autocopula models include many familiar time series as special cases. For example, it is straightforward to show that an AR(1) process, $ y_{t} = \alpha y_{t-1} + \beta + \sigma \epsilon(t)$, can be modelled using the autocopula framework using the marginal distribution $F_{\infty}(y) = \Phi\left(\frac{y-\beta/(1-\alpha)}{\sqrt{\sigma^{2}/(1-\alpha^{2})}}\right)$ (where  $\Phi$ denotes the standard normal CDF) and a Gaussian copula with mean $\mu=\beta/(1-\alpha)$ and covariance $\frac{\sigma^2}{1-\alpha^2}\begin{bmatrix}
       1 & \alpha\\
       \alpha & 1\\
       \end{bmatrix}$.

Part of the motivation for the use of autocopulas in time series modelling is that, while correlation coefficients measure the general strength of dependence, they provide no information about  how the strength of dependence may change across the distribution. For instance, in the dataset we consider here there is evidence of  \emph{tail dependence}, whereby correlation is higher near the tails of the distribution. We can quantify this using the following definition (\cite{joe1997multivariate}, Section~2.1.10).
\begin{definition}[Upper and Lower Tail Dependence]
If a bivariate copula $C$ is such that $ \lim_{u\rightarrow 1}\overline{C}(u,u)/(1-u)= \lambda_{U}$ exists, where $\overline{C}(u,u)=1 - C(1,u) - C(u,1) + C(u,u)$, then $C$ has upper tail dependence if $\lambda_{U} \in (0,1]$ and no upper tail dependence if $\lambda_{U}=0$.  Similarly, if $ \lim_{u\rightarrow 0}C(u,u)/(u)= \lambda_{L}$ exists, $C$ has lower tail dependence if $ \lambda_{L} \in (0,1]$ and no lower tail dependence if $\lambda_{L}=0$ 
\end{definition}

\begin{figure}[H]
\sidecaption[t]
    \includegraphics[width=0.62\linewidth]{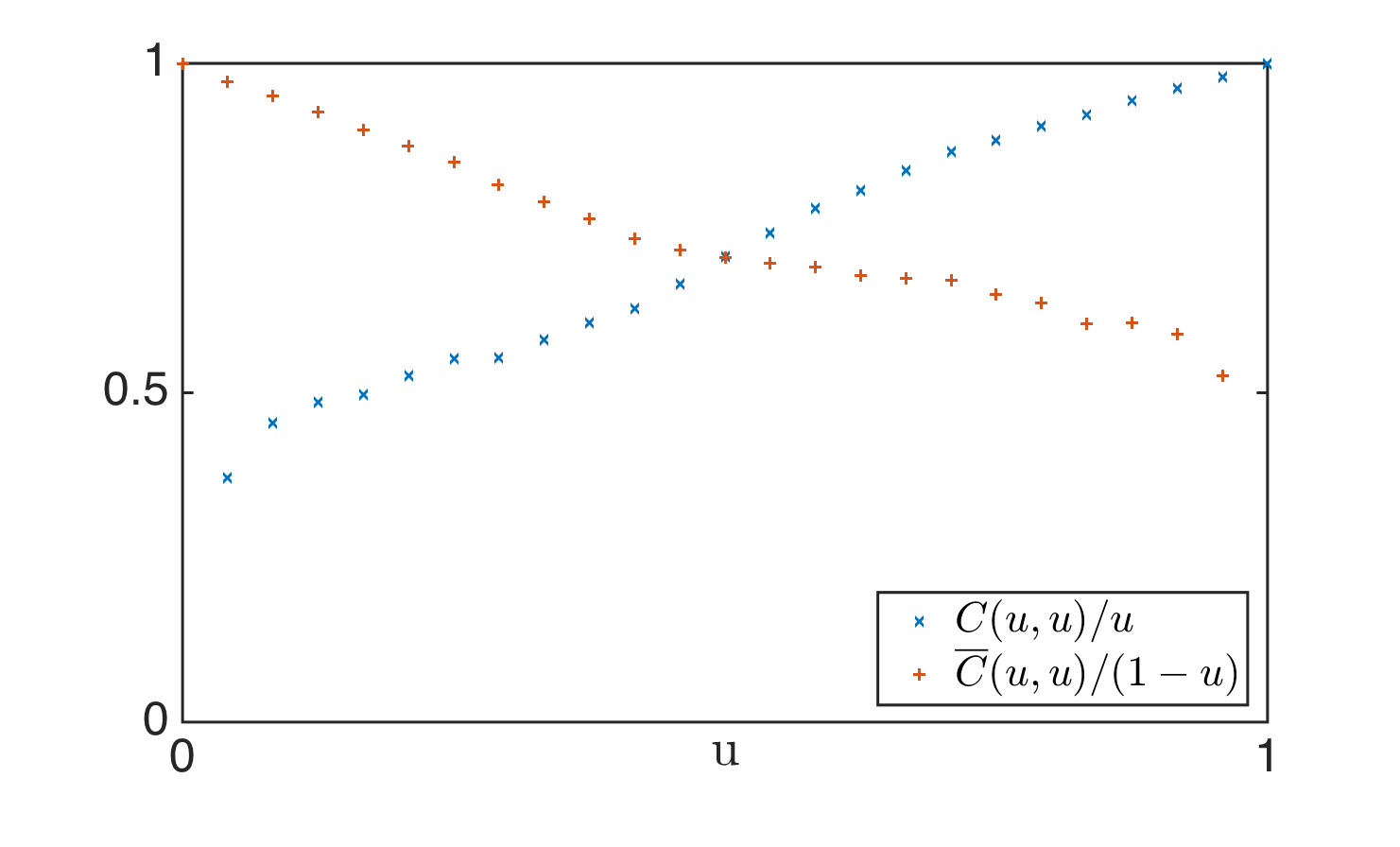}
    \caption{Estimated values of the quantities $C(u,u)/(u)$ and $\overline{C}(u,u)/(1-u)$ showing lower and upper tail dependence in the observed values of $\Delta P\Delta W$.}
     \label{fig:TailDependence}
\end{figure}

In Figure~\ref{fig:TailDependence} we show estimates of the quantities $C(u,u)/(u)$ and $\overline{C}(u,u)/(1-u)$, where here we use the order statistics of the time series $X_t=(\Delta P\Delta W)_t$ to generate a preliminary empirical proxy for the copula function $C$. It is clear from the figure that neither set of values tends towards zero in the limit $u\to 0$ or $u\to 1$, and we conclude that the data exhibit nonzero tail dependence.


\subsection{Time Varying Marginal Distribution}
\label{subsec:marginals}
As noted above, the  marginal densities for our time series will not be normal. We found that the normal inverse Gaussian (NIG) distribution provided a more satisfactory fit. More information about this distribution and its applications can be found in \cite{Barndorff-NielsenMikoschResnick2012}. Here we review its definition and properties.
 \subsubsection{Definition and properties of the NIG distribution}
 \label{subsec:4}
A non-negative random variable $Y$ has an inverse Gaussian distribution with parameters $ \alpha >0$ and $ \beta >0$ if its density function is of the form
\[
 f_{\text{IG}}(y;\alpha,\beta) =
   \frac{\alpha}{\sqrt{2 \pi \beta}}y^{-3/2} \exp{\Big(- \frac{(\alpha - \beta y)^{2}}{2 \beta y}\Big)}, \; \text{for } \; y > 0.
\]
A random variable $X$ has an NIG distribution with parameters $\alpha$, $\beta$, $\mu$ and $ \delta$ if
\[
 X|Y  =  y  \sim N(\mu + \beta y,y) \;\text{and}\;
Y  \sim \text{IG}(\delta \gamma, \gamma^{2}),
\]
with $ \gamma := \sqrt{\alpha^{2} - \beta^{2}}$, $ 0 \leq |\beta| < \alpha$ and $ \delta >0$. We then write $ X \sim \text{NIG}(\alpha,\beta, \mu, \delta)$. Denoting by $K_1$ the modified Bessel function of the second kind, the density is given by
\[
f_{\text{NIG}}(x;\alpha,\beta, \mu,\delta) = \frac{\delta \alpha \exp{\big(\delta \gamma + \beta (x-\mu)\big)}}{\pi \sqrt{\delta^{2}+(x-\mu)^{2}}} K_{1}\Big(\alpha\sqrt{\delta^{2}+(x-\mu)^{2}}\Big).
\]
There is a one-to-one map between the parameters of the NIG distribution and the  mean, variance, skewness and kurtosis of the data. We first use moment matching to determine initial estimates for the parameters; we then use these values as our initial estimates in a MLE estimation.

\begin{figure}[H]
\sidecaption[t]
    \includegraphics[width=0.62\linewidth]{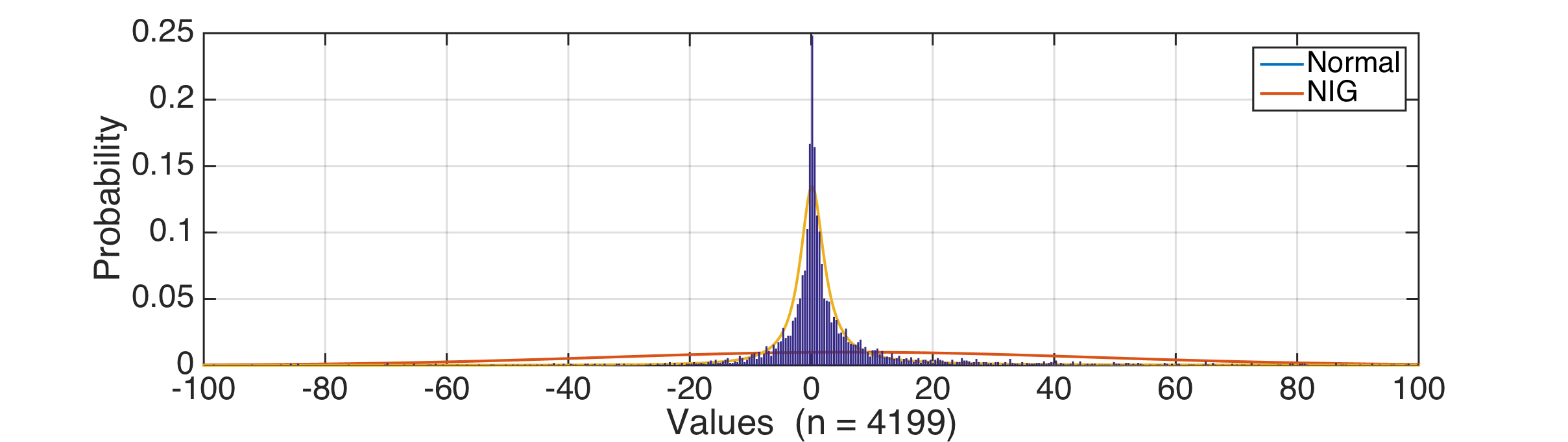}
    \caption{Histogram of observed data ($\Delta P\Delta W$), with fitted normal distribution and NIG distribution}
     \label{fig:NIGfit}
\end{figure}

\begin{figure}[H]
\sidecaption[t]
    \includegraphics[width=0.62\linewidth]{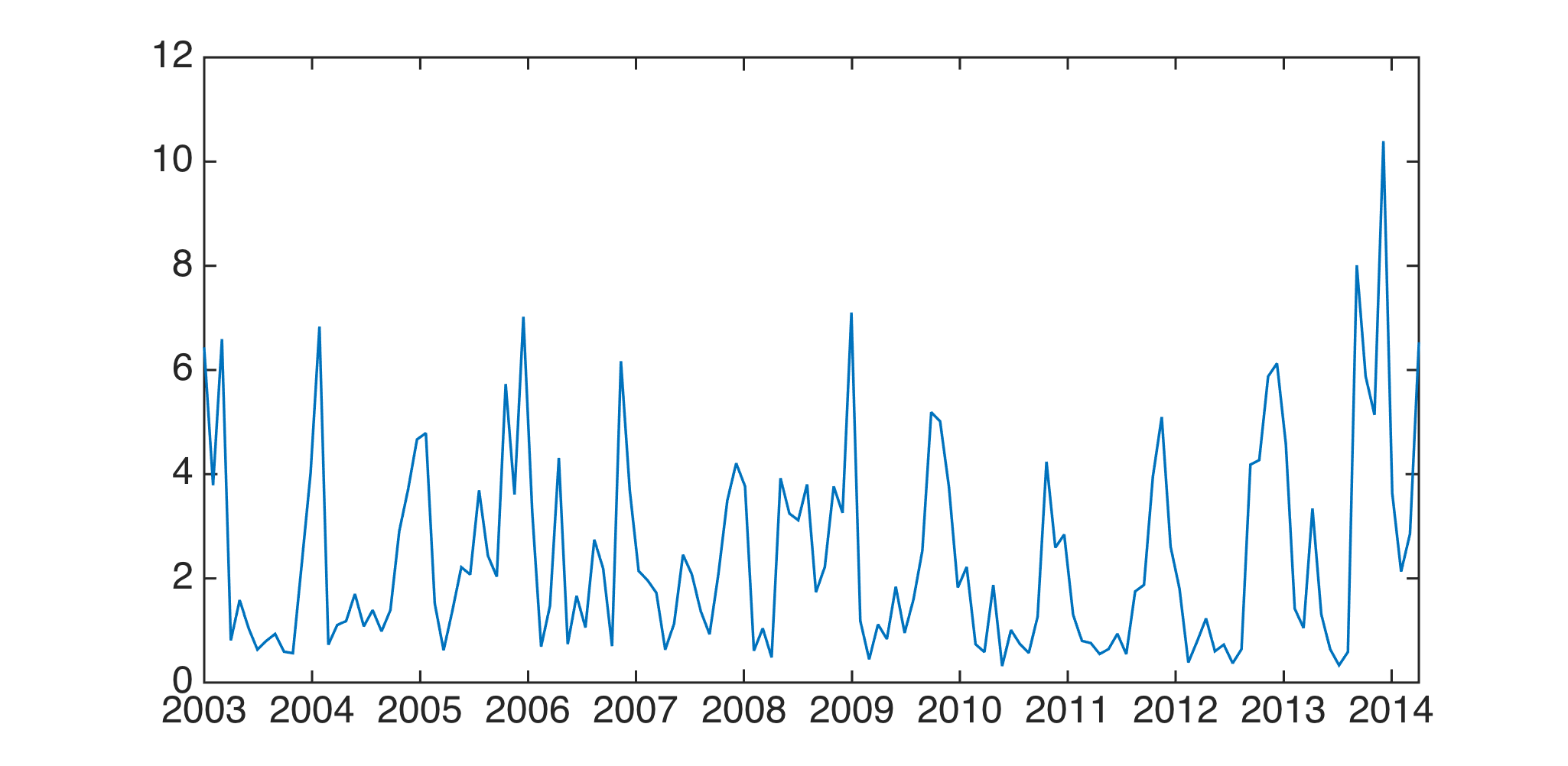}
    \caption{Calibrated monthly values of $\delta$ from the combined NIG likelihood}
     \label{fig:Delta}
\end{figure}

Table~\ref{tab:1}  shows the estimated parameters of the NIG distribution---assuming that the distribution is invariant over time. The corresponding fit to the data is shown in Figure~\ref{fig:NIGfit}, where the best fitting normal density is also shown. It can be seen that the NIG fit is quite good. However, it is evident from Figure~\ref{fig:one} that the time series is strongly seasonal. We seek to capture this seasonality through the marginal densities by making the parameter $\delta$ of the NIG distribution time-dependent. This was achieved by assuming $\delta$ to be constant in each month, and maximizing the resulting joint likelihood across the entire data set. The results are shown in Figure~\ref{fig:Delta}, and the seasonal pattern that is evident in the original data is evident again here.

\begin{table}[ht]
\caption{Results of non time-dependent NIG estimation} 
\centering
\begin{tabular}{lcccc}
  \hline
 & $\mu$ & $\alpha$ & $\beta$ & $\delta$ \\ 
  \hline
  Moment Matching & 0.3244 & 0.0231& 0.0210 &  2.7129 \\ 
\hline
  MLE & 0.0980 &  0.0131& 0.0122 &  2.3799\\ 
  \hline
\end{tabular}
\label{tab:1}
\end{table} 

As can be seen in Figure \ref{fig:Delta}, the value of $\delta$ tends to be higher in winter and lower in summer. The time series of values appears to be mean reverting  with seasonal mean and variance. We model the time series using a seasonal mean reverting process for $\nu_t = \sqrt{\delta_t}$:
\begin{equation}
\nu_{t+1} = a \nu_{t} + b(t) + \sigma(t) z_{t+1}.
\label{eq:delta}
\end{equation}
The mean and variance are estimated using periodic functions with periods from one year down to three months. 

\begin{figure}[H]
\sidecaption[t]
    \includegraphics[width=0.62\linewidth]{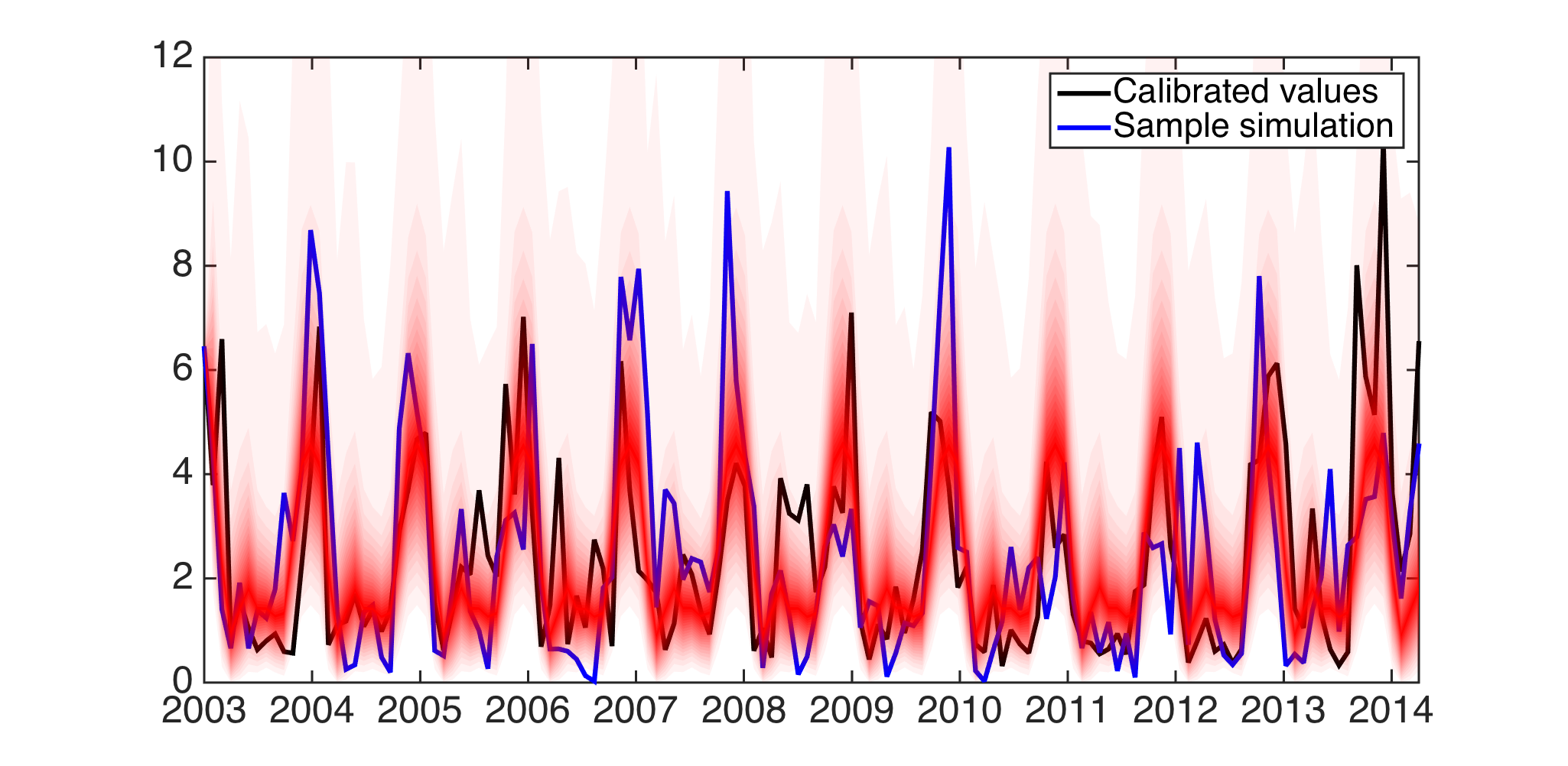}
    \caption{Calibrated monthly values of $\delta$, together with an example of a simulated path, as well as a colour contour plot of the quantiles from a large number of simulated paths.}
     \label{fig:SimulatedDelta}
\end{figure}

Simulated and estimated values of $\delta_t$ are shown in Figure \ref{fig:SimulatedDelta}. 20,000 paths were simulated using \eqref{eq:delta}, and for each month the set of values was used to determine quantiles, which were then used to create the coloured patches shown in the figure. The darker patches correspond to quantiles nearer to the centre of the distribution, and the lighter patches to quantiles nearer the extremes.

Once we have values of $\delta_t$, we can obtain the time varying cumulative distribution function and time varying density function. The NIG cumulative distribution function does not have a closed form solution, so we can compute the CDF using Gaussian quadrature to evaluate the following integral.
 \begin{equation}
 F(x_{t};\alpha,\beta,\mu,\delta) = \int_{-\infty}^{x_{t}}f_{\text{NIG}}(X_{t};\alpha,\beta,\mu,\delta_{t})dX_{t}
 \end{equation} 
In next section we explain the procedure to calculate the empirical autocopulas and simulate cash flows.

\subsection{Estimating the Empirical Autocopula}
\label{sec:AutocopulaEstimation} 
Having estimated the time-dependent NIG densities, we use these to produce a time series of values $V_t = F_t(X_t)\in[0,1]$. If the marginal densities were exact, these would be uniformly distributed on $[0,1]$. In practice, they will only be approximately uniform, and we generate an additional empirical marginal density and an empirical (auto)copula to capture the joint density of $(V_t,V_{t-1})$.

The empirical autocopula $C$ is estimated by first estimating an empirical joint density for $(V_t,V_{t-1})$ in the form of a strictly increasing continuous function $\Phi(\cdot,\cdot)$ that is piecewise bilinear. The domain $[0,1]^2$ is partitioned into rectangles containing approximately similar numbers of samples $(V_{t-1},V_t)$, and taking $\Phi$ to be the cumulative integral of the sum of indicator functions for these rectangles, scaled by the number of samples in each rectangle. $\Phi$ is then used to create strictly increasing piecewise linear marginal densities $\Phi_{1}$ and $\Phi_{2}$. The inverses of these densities are therefore also piecewise linear, and when composed with $\Phi$ they generate a piecewise bilinear copula function $C(u_1,u_2) = \Phi\big(\Phi_{1}^{-1}(u_1),\Phi_{2}^{-1}(u_2)\big)$.

\begin{figure}[H]
    \centering
    \begin{subfigure}[b]{0.31\textwidth}
        \includegraphics[width=\textwidth]{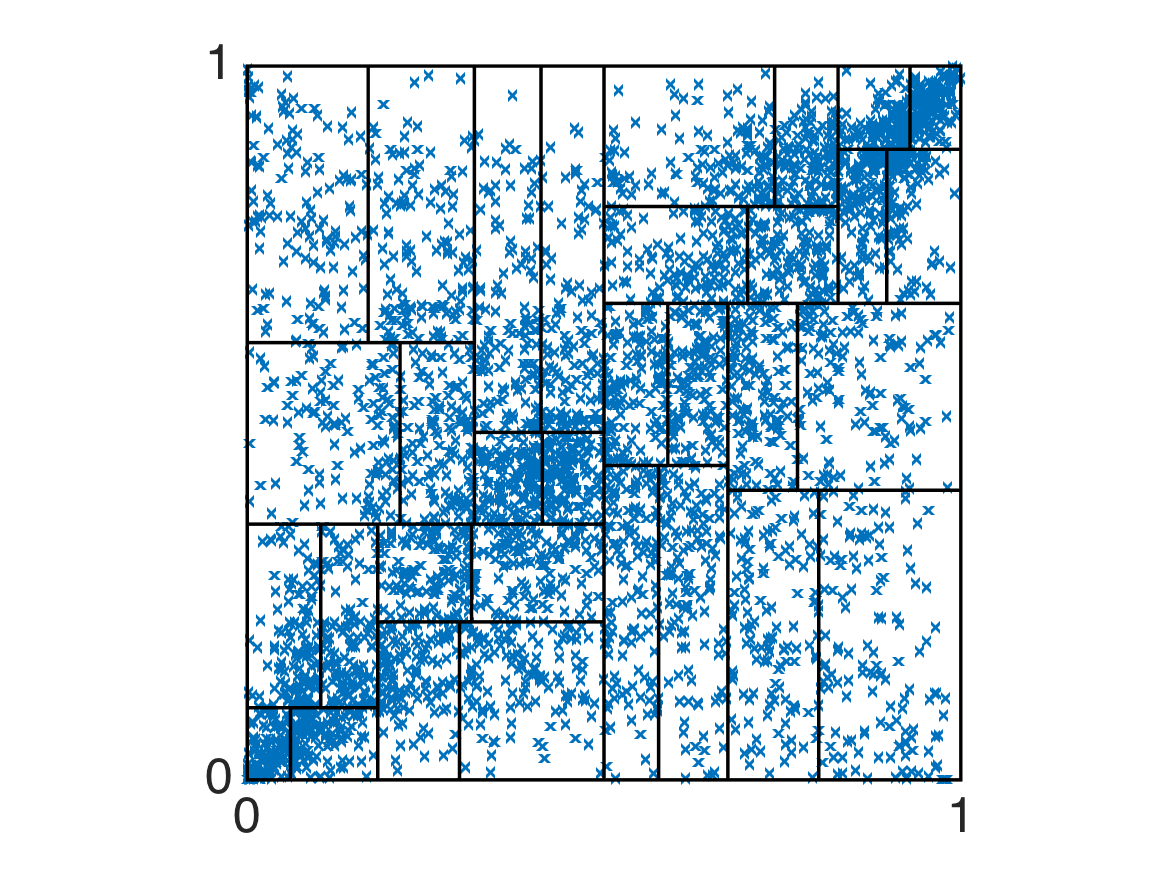}
        \caption{Scatter plot of $\Phi_{1}(V_{t-1})$ against $\Phi_{2}(V_t)$. Each rectangle contains about the same number of points.}
        \label{fig:Scatter}
    \end{subfigure}
    \hfill
    \begin{subfigure}[b]{0.31\textwidth}
        \includegraphics[width=\textwidth]{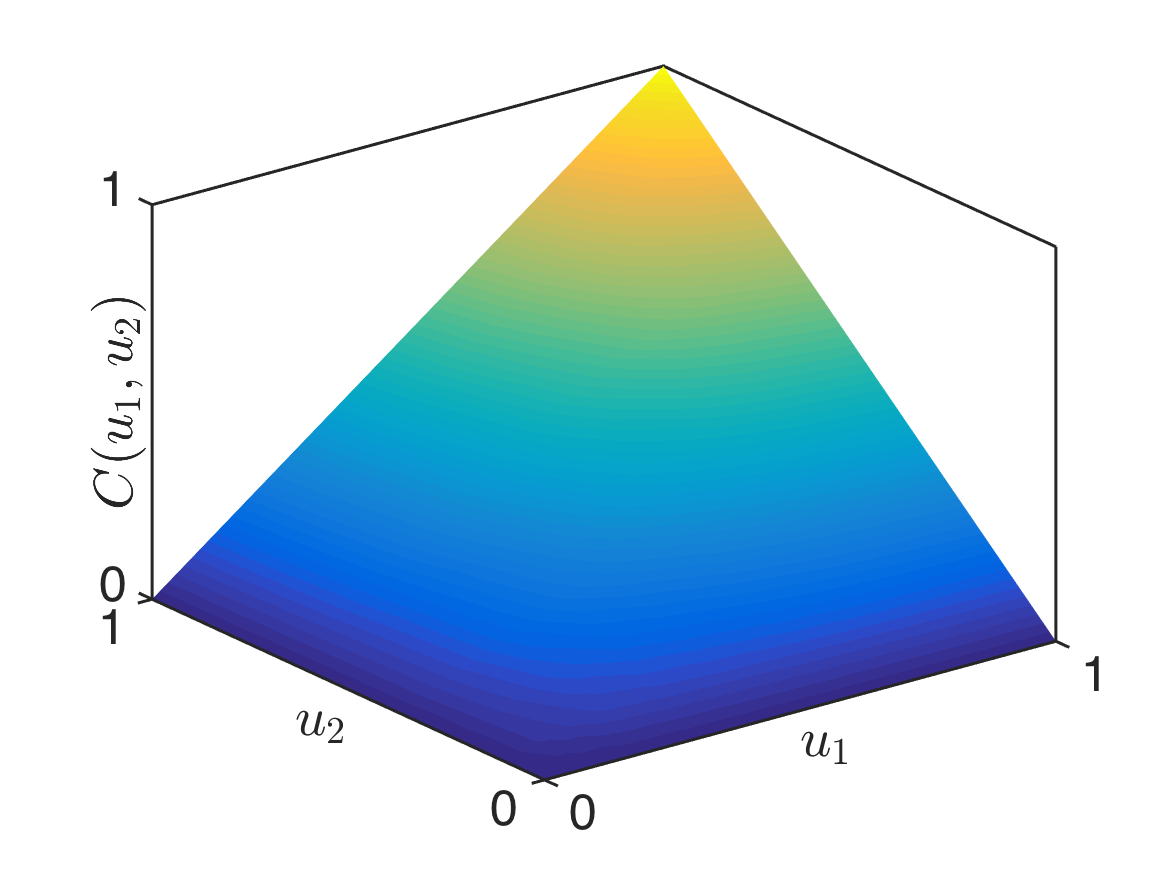}
        \caption{Empirical autocopula $C(u_1,u_2)$ defined to be bilinear on each of the rectangles shown in (a).}
        \label{fig:copula}
    \end{subfigure}
    \hfill
    \begin{subfigure}[b]{0.31\textwidth}
        \centerline{\includegraphics[width=\textwidth]{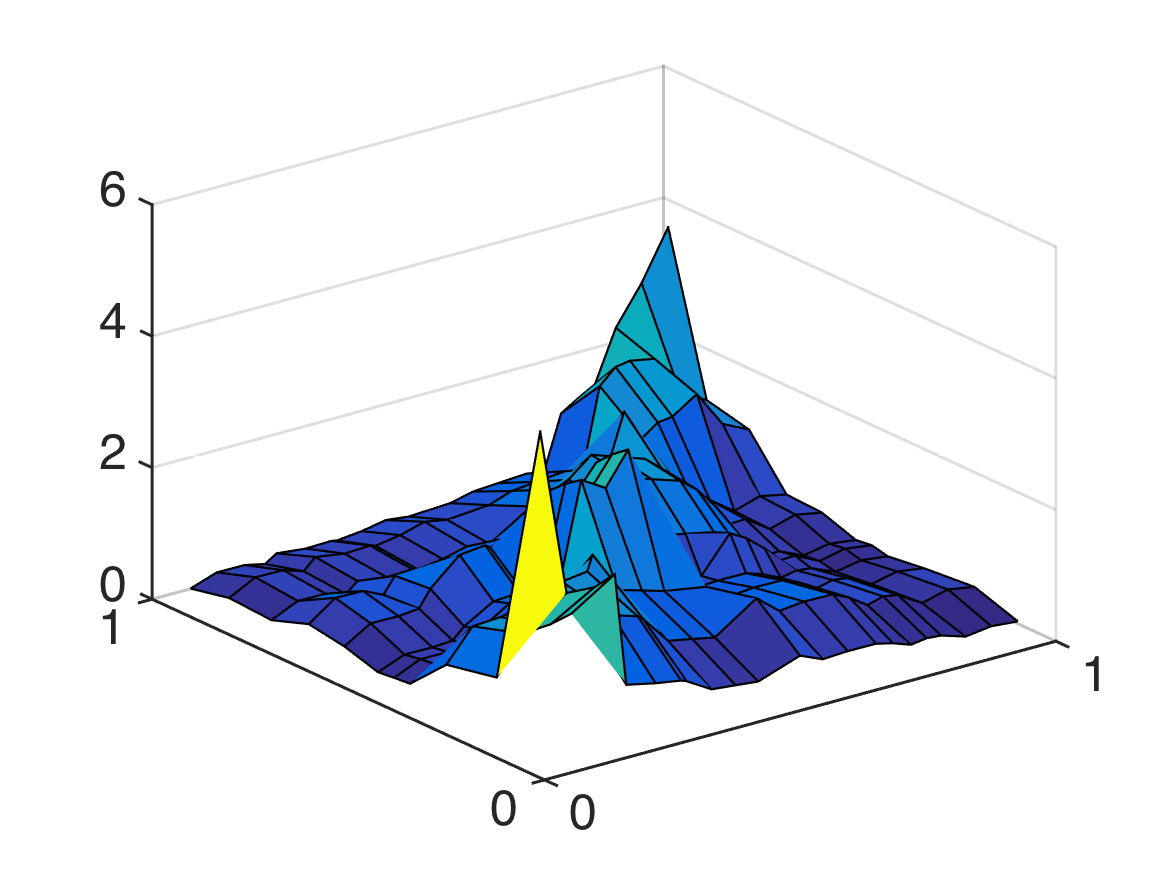}}
        \caption{The empirical density $\partial^2C/\partial u_1\partial u_2$, which is constant on each of the rectangles shown in (a).}
        \label{fig:density}
    \end{subfigure}
    \caption{Generation of the empirical autocopula}
    \label{fig:empiricalautocopula}
\end{figure}

This process is illustrated in Figure~\ref{fig:empiricalautocopula}. In Figure~\ref{fig:Scatter} we plot the pairs of transformed values $\big( \Phi_{1}(V_{t-1}),\Phi_{2}(V_t)\big)$, together with the outlines of rectangles used to generate the piecewise bilinear function $C$. As mentioned, these rectangles contain roughly equal numbers of points; constructing the empirical autocopula in this way ensures that it is strictly increasing, and well-suited to enable the computations involved in time series simulation (see below) to be carried out efficiently. 

The resulting empirical autocopula $C$ is shown in Figure~\ref{fig:copula}. This function is binlinear on each of the rectangles shown in Figure~\ref{fig:Scatter}, but is less regular than it looks. The corresponding joint density, $\frac{\partial^2C}{\partial u_1\partial u_2}(u_1,u_2)$, is shown in Figure~\ref{fig:density}. It can be seen that the density is higher near $(0,0)$ and near $(1,1)$, which is consistent with the tail dependency observed earlier.




\subsection{Simulation of time series using autocopula }
\label{subsec:simulation}
Armed with the time-dependent NIG densities $F_t(\cdot)$, the empirical marginal densities $F_{V,i}(\cdot)$ and the empirical autocopula $C(\cdot,\cdot)$, we can generate simulated values $x_t$ as follows.
\begin{enumerate}
\item Given an initial value $x_0$, generate $v_0=F_0(x_0)$.
\item For $t=0,1,\dots$, given $v_t$, generate $v_{t+1}$:
  \begin{enumerate}
  \item Set $u_1 = \Phi_{1}(v_t)$.
  \item Given $u_1$, create the piecewise linear function $\underline{C}(u):=C(u_1,u)/u_1$.
  \item Set $u_2 = \underline{C}^{-1}(U)$, where $U$ is an independent uniform random draw.
  \item Set $v_{t+1} = \Phi_{2}^{-1}(u_2)$.
  \end{enumerate}
\item For each $t>0$, set $x_t = F^{-1}_t(v_t)$.
\end{enumerate}
Here we have used the fact (already alluded to in \eqref{eq:conditionalCDF}) that, if $U_1$ and $U_2$ are uniform random variables whose joint distribution is the copula $C(u_1,u_2)$, then, for $u_1>0$, the cumulative density function for $U_2$, conditional on $U_1=u_1$, is
\[ P[U_2<u_2 | U_1=u_1] = \frac{\partial C}{\partial u_2}(u_1,u_2) = \frac{C(u_1,u_2)}{u_1}. \]
The proof of this can be found in, for example, 
 \cite{darsow1992copulas}.

The fact that $C$ is a piecewise bilinear function means that $\underline{C}$ will be piecewise linear. Moreover, the construction of the empirical copula as described in Section~\ref{sec:AutocopulaEstimation} ensures that it is an increasing function with a limited number of corners. Its inverse can then be constructed readily, and will also be an increasing piecewise linear function with a limited number of corners, and so can be evaluated with little computational effort. Indeed, in practice the computation of the final step in the above algorithm, the inversion of the time-dependent NIG densities, took more time than the copula-related computations.

\section{Results}
\begin{figure}[H]
\sidecaption[t]
    \includegraphics[width=0.62\linewidth]{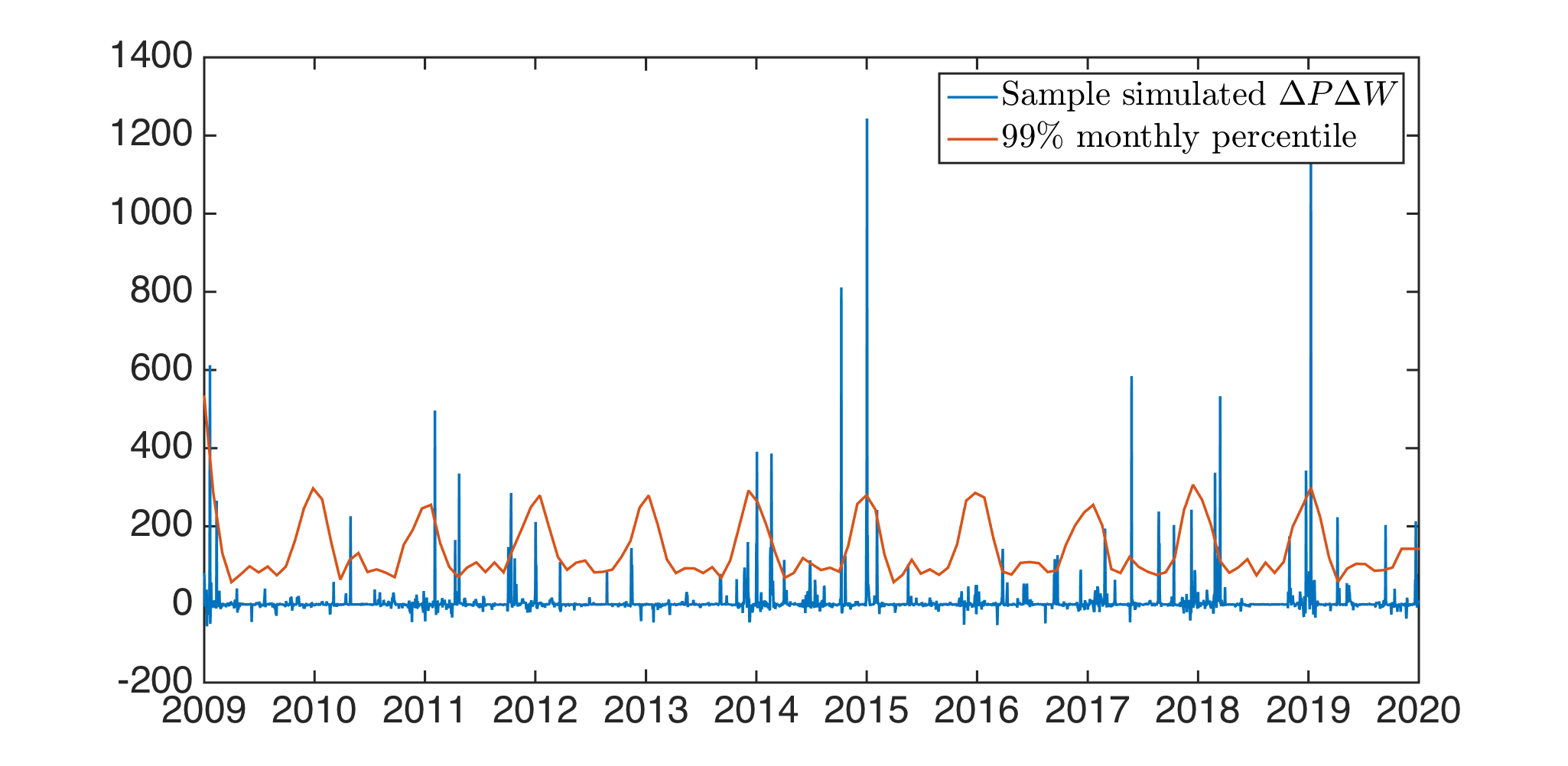}
    \caption{Simulated values of $\Delta P\Delta W$, together with the 99th percentile of collected monthly values from around 700 simulations.}
     \label{fig:SimulatedCashFlows}
\end{figure}

In Figure~\ref{fig:SimulatedCashFlows} we show a 12-year sample time series  for $\Delta P\Delta W$ computed as described in Section~\ref{subsec:simulation}. In addition, we simulated around 700 independent time series and computed, for each month, the 99th percentile of values produced in that month across all simulations. 

What can be seen in the sample path is the same mixture of quiescent periods and periods with extremely large deviations from zero. There is some evidence of `clumps' of large deviations occuring in winter months, although this is less clear than in the original data (see Figure~\ref{fig:one}). There is, nevertheless, an increased occurence of large deviations in winter months, as can be seen from the plot of the 99th percentiles that is superimposed on the sample simulation shown in Figure~\ref{fig:SimulatedCashFlows}.

\begin{figure}[H]
\sidecaption[t]
    \includegraphics[width=0.62\linewidth]{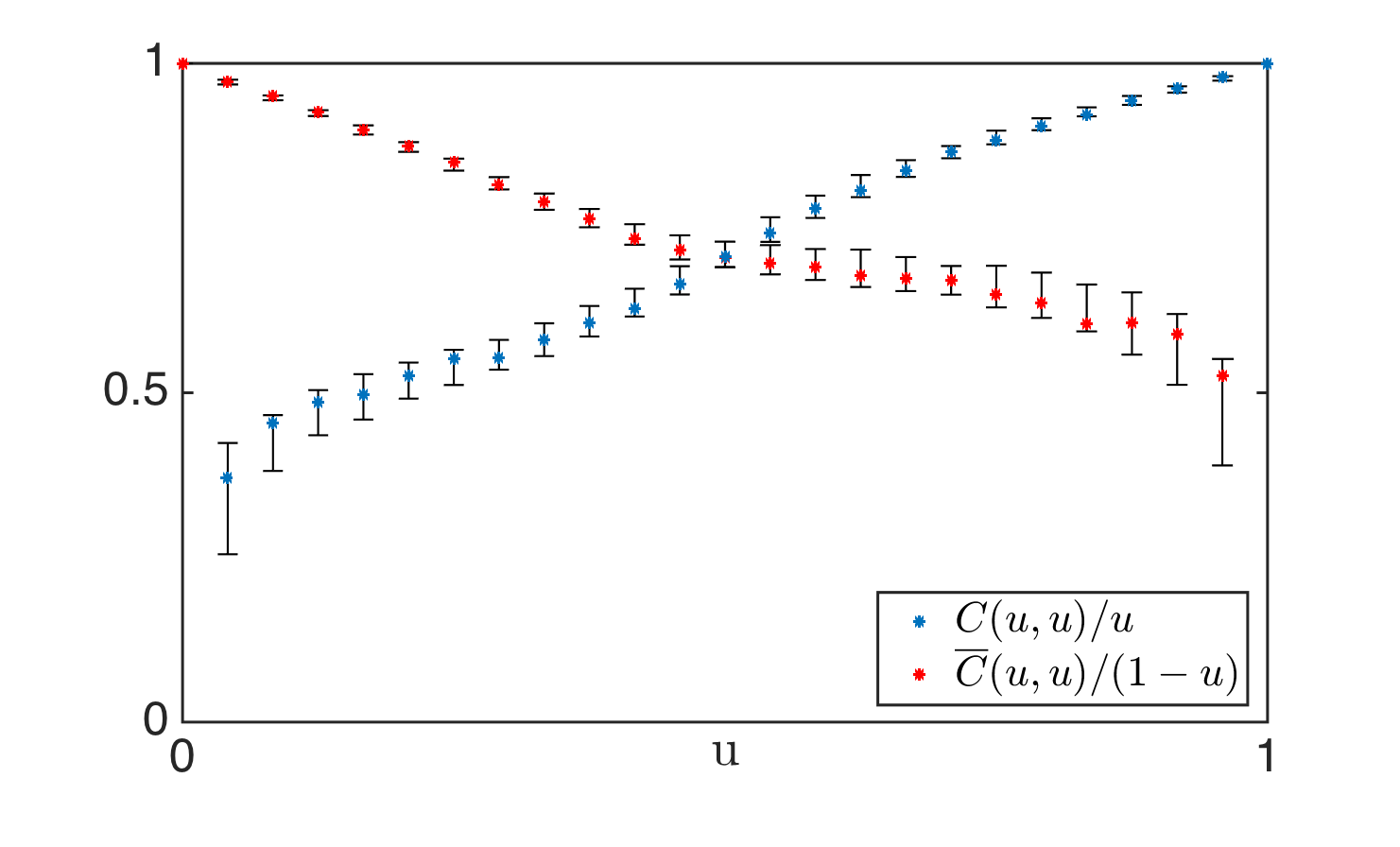}
    \caption{Estimated values of the quantities $C(u,u)/(u)$ and $\overline{C}(u,u)/(1-u)$ for the original observations of $\Delta P\Delta W$. Also shown are error bars corresponding to the 5th and 95th percentiles of the values obtained from around 700 simulations.}
     \label{fig:TailDependenceErrorBars}
\end{figure}

In Figure~\ref{fig:TailDependenceErrorBars} we illustrate the fact that the simulations have reproduced the tail dependence that was evident in the time series of original observations. The data from Figure~\ref{fig:TailDependence} is reproduced, together with error bars corresponding to the 5th and 95th percentiles of the values obtained from the simulations.

\bibliographystyle{abbrvnat}
\bibliography{bib}

\end{document}